# Voice-Based Conversational Agents for self-reporting fluid consumption and sleep quality

Abdalsalam Almzayyen, Ángel Vela de la Garza Evia, Nick Coronato, Mehdi Boukhechba

**Abstract**—Intelligent conversational agents and virtual assistants, such as chatbots and voice assistants, have the potential of augmenting health service capacity to screen symptoms and deliver healthcare interventions. In this paper, we developed voice-based conversational agents (VCAs) in the Google Actions Console to deliver periodic self-assessment health surveys. The focus of this paper is to accommodate self-monitoring for patients with specific fluid consumption requirements or sleep disorders. Our VCAs, named FluidMonitor and Sleepy, have been tested to integrate naturally into a patient's daily lifestyle for the purpose of providing useful interventions. We show the functionality of our Google Actions and discuss the considerations for using VCAs as an at-home self-reporting survey technique. User testing showed satisfaction with the ease of use, likeability, and burden level of the VCAs.

**Index Terms**—Conversational Agents, Voice-Based Conversational Agents, Fluid Consumption Monitor, Sleep Diary

✦

## 1 Introduction

Voice-based conversational agents (VCAs) use artificial intelligence and natural language processing algorithms to interact and converse with humans. The most popular commercial VCAs are Amazon's Alexa, Google's Google Assistant, and Apple's Siri. These can be found on devices like smartphones, smart speakers, home appliances, and cars. As a result, VCAs' network expands to over 2.5 billion devices globally [1]. With the rising popularity of affordable smart speakers like Amazon Echo or Google Home, VCAs have become part of the day-to-day routine of many individuals. In 2020, it was estimated that around 90 million U.S. adults owned a smart speaker [2]. Due to their widespread availability and adoption, researchers and healthcare providers are working on understanding how VCAs in smart speakers can contribute to the healthcare space. Some of the current general uses include asking for information about symptoms or medications, providing directions to the nearest hospital, scheduling medical appointments, or finding a healthcare provider in the area [1]. In addition to these, VCAs are also being used for self-reporting health surveys. The data collected from these surveys can provide valuable insights into a patient's health. Traditionally, these surveys are administered by pen and paper or through a smartphone application, but by leveraging VCAs' natural language processing power and infrastructure, they can now be administered through voice.

There are multiple health conditions that can benefit from self-reported health data. For example, pregnancy, breast-feeding, and bladder infections require patients to reach a minimum threshold of fluid intake in a particular time period [3]. Other conditions, such as kidney disease, require patients to limit their daily consumption of fluids in order to meet the demands of dialysis or other such procedures [4]. Accurately self-reporting daily log of fluid consumption would be beneficial for these medical conditions. Another example is tracking sleep quality and quantity for the monitoring of many physical and behavioral health conditions. Patients diagnosed with insomnia or other sleep disorders keep track of their sleeping patterns through self-reported surveys such as a sleep diary. The purpose of this project is to develop VCAs for these particular self-reporting cases of fluid consumption and sleep monitoring.

## 2 Related Work

[5] performed a systematic literature review on the use of VCAs to prevent and manage varying health conditions. Three of the twelve studies specifically focused on VCAs on smart speakers. In [6], a VCA for Google Home was created to help people with disabilities develop and practice their social skills. The VCA walked the participants through branching storylines that involved social interactions like asking a store employee a question. At each stopping point the participants chose one of three options labeled as good, adequate, and negative. [7] built a VCA for Google Home called Healthy Coping with Diabetes to provide elderly patients diagnosed with type 2 diabetes a better tool for tracking and managing their health condition. Some of its functionalities include administering PHQ-9 depression and monitoring surveys, providing personalized feedback from the logged survey data on metrics like glucose level, and answering dietary questions to inform whether a food item is good to eat or not. As part of their proposed system's pipeline, collected data will get extracted and published on a website for the patients to access and visualize. Researchers from this study also conducted a satisfaction survey on ten elderly participants. Results showed that the majority of the participants would use the VCA over a smartphone application and that they liked the naturalness of the interaction with the system. [8] measured the speech recognition threshold (SRT) from self-administered listening tests on an Alexa smart speaker. They evaluated the impact that differing hearing loss levels and room acoustics had on the test's reliability. Results showed that smart speakers can be used to administer listening tests and that data collected from these tests can detect deviations in the SRT.

Other research has focused on the design of VCAs for self-reporting health data. [9] conducted a study with 59

participants to understand how responses to the WHO-5 well-being survey differed when answering on paper and through a VCA on Google Home. Researchers also analyzed the impact that discrete or open-ended response requirements had on the overall user experience and the time it took to complete the interaction with the VCA. After completing the WHO-5 questionnaire on paper and the discrete or open-ended versions of the VCA, participants answered the Subjective Assessment of Speech-system Interface Usability (SASSI) questionnaire that measures how well a VCA performs in areas like accuracy, cognitive demand, likability, and speed. Results from this study indicated that the paper and VCA re-sponses to the WHO-5 survey match up and that the response format (discrete vs. open-ended) does impact users' overall experience with the VCA. For example, it took participants more time to do the WHO-5 on the discrete version of the VCA because they would forget the response scale or give an invalid answer choice. However, participants noted that the opend-ended version of the VCA introduced uncertainty since they did not know what to respond. Overall, [9]'s work demonstrated that VCAs can be used for self-reporting health data and that it is important to consider the system design in the development stage.

These studies also highlighted the challenges present in VCAs for healthcare. [6] worked with a population that had speech impairments. The Google Assistant incorrectly parsed most of the responses to the story scenarios. [7] noted how the speed of the VCA narration could impact the user's experience with the system. Furthermore, since they were working with an elderly population, the VCA was not suited for those with hearing loss. [8] described that further research was required to assess how actual room acoustics, like in a user's home, could influence the effectiveness of the self-administered listening test. Finally, [9] mentioned the difficulties in mapping open-ended question responses to a fixed scale. Despite these challenges, the development of accessible and affordable VCAs present the opportunity for people to self-report their health data from the convenience of their homes.

## 3 Methods

By creating an end-to-end open-source application for Google Assistant, we address two medical cases where self-reported surveys could benefit the user (patient) as well as their caregivers or medical providers. In section 3.1, we go through the technical details of the development of our VCAs. Afterwards, in section 3.2, we discuss the development of FluidMoni-tor, aimed at improving the process for daily recording of fluid consumption. Finally, section 3.3 discusses the methods and outcomes of Sleepy, a separate voice-based survey that captures sleep quality and quantity information for further analysis. The focus of this research is to provide a proof of concept for VCAs to assist with medical self-reporting requirements. The problem was divided into four nested components, as shown in Figure 1.

Ultimately, engineering behavior change for patients' well-being can be framed as a behavioral design problem. Behavioral design can be defined as the practice of changing things from where they are to where we want them to be, considering the environmental and behavioral science components of the system. To facilitate behavioral change in the medical setting, [10] recommends that we "focus on the management of behaviors that support or diminish health." In the 21st century, we should leverage digital technologies to test behavioral theories in "real-world" contexts using behavior change techniques. In framing our problem as a behavioral design approach, our goal was to engineer behavior change that supports patients' well-being across a variety of health scenarios.

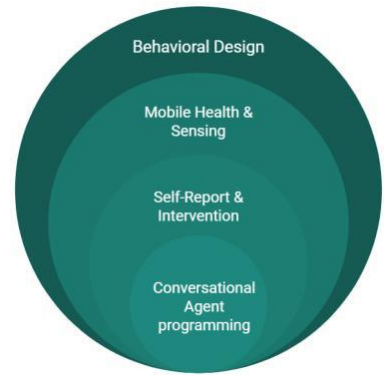

Fig. 1. The design challenge for our research.

The specific field which enables this goal is Mobile Sensing and Health. Our research applies current mobile technology to enable patients to easily record important health data. As a rapidly developing field, mHealth has received an abundance of recent attention. According to [11], "technology can provide detailed, unobtrusive assessment of behavior and its context, while complementary qualitative methods are crucial to fully understand and interpret user experiences." The second goal of our study was to sense the relevant data, analyze it, and provide context-appropriate interventions.

We chose to accomplish these tasks by examining the util-ity of self-reported surveys that can be delivered to patients outside the clinical setting. Self-reported survey responses can help the user to accurately track their health status without a great deal of burden. By understanding the best context in which to deliver self-reported surveys, we optimize the data's accuracy and overcome some of the biases that are inherent in this kind of approach (i.e. self-reporting bias and recency bias).

The technical method for capturing the data was to program a VCA that is capable of periodically querying the patient with verbal questionnaires. This is the final goal of our research, and our output is a proof of concept.

### 3.1 Technical Implementation

Our implementation utilized the Google Actions Console, and it provided the recognition of intentions and the direction of the possible conversation flows for each survey. The Google Actions Console was useful for design and simulation of simple functions, as it has an intuitive and generally code-free interface for development, testing, and deployment. For more complex functions, we implemented webhooks that trigger at various points in conversation to securely store and retrieve information in the database, which enable more functionalities to call specific variables of interest, such as total cups drank, for a more personalized user experience.

We used Firestore for the database, and because it was on the Google Cloud platform we were able to simply connect to the database. Data logs are created at every user response to our VCAs FluidMonitor and Sleepy, as well as on any changes (to keep track of cups drank, for example). Each iteration of our software was rolled out to Alpha users for testing, and Google's deployment workflow allowed for simple version control. Additionally, the Google Actions Console provides a useful analytics tab for tracking app usage and user engagement over time. Our final VCA applications can be downloaded directly to a user's smartphone through the Google Assistant app or Google Home device. The survey tools can be programmed in such a way that the interactions between user and VCA are seamless and efficient for daily use.

The data analysis portion consisted of exporting the data in Firestore, filtering out any unneeded entries, and then linking those to a data analysis platform in order to analyze and display the data. To be able to provide interventions we needed to use Google Account linking to connect the user's account to our database and manipulate their data. This required the implementation of an authorization server for two-way authentication between the user and our service. We were able to authenticate users using a token sent to our service from the user. After successful authentication, a message is sent back to the user to input their password; after confirming sign-in, the user is recognized. This allows for improved interactions because when the user is recognized, there is no need to confirm the user's Participant ID during the survey.

Linking users in the database also enables the future potential for storage and retrieval of user-specific variables. For example, the daily goal for each individual is currently hard-coded at the beginning of their experience with Fluid-Monitor; ideally, this consumption goal should be dynamic in nature. It would be useful to account for other individualized aspects (such as daily physical activity or the previous day's consumption) to automatically adjust the dynamic health goal. By capturing external user-specific information across conversations, our Google Action could allow for adjustment of the consumption goal or limitation through automated calculations generated by artificial intelligence.

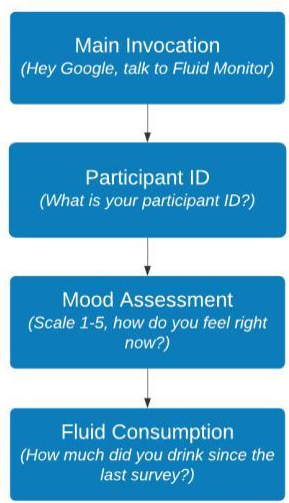

Fig. 2. Simplified diagram of FluidMonitor conversational flow

### 3.2 Fluid Consumption Monitor

We developed FluidMonitor, a Google Action that is intended to facilitate multiple daily self-surveys for either the fluid-limited or fluid deficient patient. The Google Action can be used with any interface that is Google Assistant-enabled, to include smart home speakers (with or without touchscreen monitor), smartphones, and smart headphones. Our test experiments were restricted to two configurations: the Google Nest Mini smart speaker and the Google Assistant app on iOS operating system. The Google Nest Mini is a voice-only device, while the Google Assistant app allows for voice as well as touchscreen input.

The timing schedule for which FluidMonitor is best implemented is still an open research question; for this study, we chose to invoke FluidMonitor each day at 9:00am, 3:00pm, and 9:00pm. Figure 2 illustrates a simplified diagram of our VCA conversational flow. The user initiates the action with a verbal invocation: "Hey Google, talk to Fluid Monitor." After this, the user is prompted to answer three short questions: confirm their User ID, note their current health status on a 1-5 scale, and recall their fluid intake since the last survey. This whole process takes less than a minute to complete.

FluidMonitor has the capability to record each survey response in Google Firestore, storing the verbal responses as either conversation parameters (within conversations) or user parameters (across conversations). If FluidMonitor fails to understand the user or does not hear a response within 10 seconds, it asks the user to repeat themselves before continuing or ending the conversation.

FluidMonitor can then perform simple calculations to determine whether the user has met or exceeded their daily fluid goal or limit, though this feature was not tested as part of this research endeavor. Those with access to the database can retrieve a variety of useful reports, such as a plot of the longitudinal fluid consumption rate during any particular time period. Through automated or manual analysis of this data, effective interventions can be programmed to guide the patient towards meeting their consumption goals.

### 3.3 Sleep Survey Diary

Our second contribution was Sleepy, which is based on previous research for the SHUTi program (Sleep Healthy Using The Internet) [12], [13]. SHUTi is a fully automated online cognitive behavioral therapy tool for assisting individuals suffering from insomnia-type sleep disorders. The SHUTi sleep diary questionnaire, which we used for our VCA, is constructed as a text-based diary for tracking sleep events, the quality of sleep, external factors, and general notes on sleep (see Appendix A). Once per day, Sleepy is delivered to participants via voice-based conversation in order to record 11 pieces of data. Like FluidMonitor, Sleepy can be accessed on any Google Assistant device or smartphone. Sleepy records survey responses securely in a Firestore database (separate from FluidMonitor). These responses can then be exported in a variety of formats to the user. For example, nightly sleep quantity can be plotted in a dashboard-type output for the patient to review their sleep patterns. With further work, Sleepy has the capability to suggest interventions to the patient in order to assist with their sleep cycle management.

TABLE 1
Evaluation Survey Questionnaire Results for FluidMonitor
(1 = Strongly Disagree, 3 = Neutral, 5 = Strongly Agree )

| Question | Avg. | St. Dev. |
| --- | --- | --- |
| EASE OF USE: The system was very easy to use. | 4.67 | 0.47 |
| NATURALNESS: The system felt very natural. | 3.89 | 0.87 |
| ACCURACY: I am very confident that the system captured my responses correctly. | 4.56 | 0.50 |
| PRIVACY: I am very confident that the system is not storing any unnecessary personal health data. | 4.22 | 0.92 |

TABLE 2
Evaluation Survey Questionnaire Results for Sleep Survey
(1 = Strongly Disagree, 3 = Neutral, 5 = Strongly Agree )

| Questions | Avg. | | | St. Dev. | | |
| --- | --- | --- | --- | --- | --- | --- |
| | Pen and Paper | Online Survey | Google Action | Pen and Paper | Online Survey | Google Action |
| The system is accurate | 4.42 | 4.17 | 3.92 | 0.51 | 0.58 | 0.79 |
| The system didn't always do what I wanted. | 2.17 | 2.42 | 2.67 | 1.19 | 1.00 | 1.07 |
| The interaction with the system is efficient | 3.92 | 3.58 | 4.33 | 0.90 | 1.16 | 0.65 |
| The system is friendly | 4.08 | 3.67 | 4.50 | 0.51 | 1.23 | 0.52 |
| I would use this system. | 3.83 | 3.33 | 4.08 | 0.94 | 1.15 | 0.67 |
| The interaction with the system is boring | 2.92 | 2.67 | 2.17 | 0.90 | 1.07 | 1.03 |
| I always knew what to say to the system | 3.92 | 4.00 | 4.08 | 1.08 | 0.95 | 0.79 |
| The interaction with the system is fast | 3.42 | 3.67 | 4.00 | 1.38 | 1.15 | 0.85 |

## 4 Discussion

### 4.1 Assessing the User Experience

User experience testing was fundamental in the iterative development process for both of our Google Actions. An important theme for this research was to understand and validate how VCAs can be used for positively affecting the lifestyle habits of patients without increasing their burden. Feedback from real users allowed us to make improvements to the existing VCA software, with the end goal of deploying Google Actions that are unobtrusive and suitable for daily use. To assess usability of FluidMonitor, we conducted a test of 11 separate users. Seven users tested FluidMonitor on the Google Nest Mini speaker, and four performed the test on an iOS smartphone. Following their interaction with FluidMonitor, we asked our testers to evaluate four categories: ease of use, naturalness, accuracy, and privacy. Their responses were scored on a scale of one to five, with five being the most positive response. The results of our test are presented in Table 1.

Feedback indicated that FluidMonitor was very easy to use and we assessed that users were confident that the system accurately captured their voice responses. The low mean score for naturalness (3.89 0.87) led us to improve some of the conversation components of the VCA.

To evaluate Sleepy, we asked 12 prospective users to complete the sleep diary in three formats: written (pen and paper), online (Google Form survey), and with Sleepy on the Google Nest Mini smart speaker. We recorded the time required for users to complete each format. Then, users answered eight qualitative questions from the SASSI questionnaire [14]. These questions helped us to gather data for system response accuracy, likeability, annoyance, and speed. Again, the score of five represents "Strongly Agree." Amongst the three options, the Google Action scored highest in six of eight SASSI questions. Our user base felt that a written survey can most accurately capture their responses, which we expected; there is no time limit for users to consider when writing the SHUTi diary responses and there is no doubt about what was recorded in each row. The pen and paper method also scored highest with regards to always doing what the user wanted; again, the user can access and review all of their responses for extra consideration or to make corrections. Sleepy, on the other hand, depends on the VCA hearing and understanding the speech response within a finite period of time (about 10 seconds). When using Sleepy on a smartphone or smart speaker with display, the user has the advantage of seeing their responses in real time as recorded by the VCA. The results of our evaluation for the three methods are shown in Table 2.

We also measured burdensome levels, format preference, and time for each format. None of our 12 testers reported that Sleepy was "Burdensome" or "Really Burdensome." When asked for format preference, two-thirds of participants indicated that they would use Sleepy over the other formats.

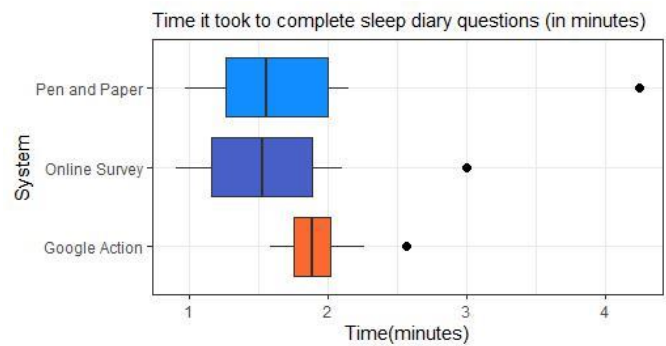

Fig. 3. Time to complete sleep diary questionnaire by format

As shown in Figure 3, the median survey completion time was highest in Sleepy, requiring just under two minutes. However, we note that total time is not the only factor to consider; both the written and online versions of the sleep diary require additional time and burden to initiate (i.e.

finding a pen and diary or logging into a web portal). These methods cannot be easily multitasked or seamlessly integrated into the user's daily habits, as can be done with a VCA.

### 4.2 Alpha User Data Collection

The second part of the project evaluation consisted of Alpha user testing. The purpose was to assess initial user adherence rates (rate at which users actually completed their assigned VCA surveys) as well as to glimpse the effectiveness of our Google Actions as behavior-change tools. Three users were selected for this validation; 57 user-days of data were collected for FluidMonitor and 21 user-days for Sleepy. After extracting this data from Firestore, we demonstrated the type of output plots which are made possible by FluidMonitor and Sleepy. Figure 4 below shows a sample visualization of daily fluid consumption with real data from our Alpha user base.

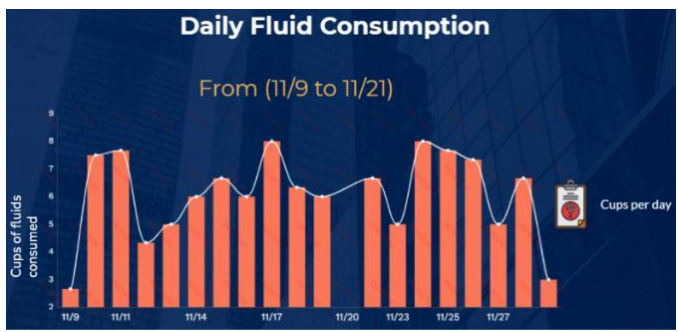

Fig. 4. Sample visualization of FluidMonitor data (mean fluid consumption for three Alpha users).

Initial results indicate that our test users did adhere to the scheduled VCA survey interactions on a daily basis. This also validated that data extraction and visualization provide useful feedback to an interested patient or healthcare provider. From this kind of plot, users or clinicians can make informed decisions regarding future sleep or fluid consumption protocols.

We propose that some level of reactive behavior change is made possible just through the process of conducting daily diaries; in other words, the simple act of self-monitoring in this way can lead to a beneficial result by offering support for committing to the intervention. This phenomenon has been noted in previous research [15], [16] as an evidence-based strategy to encourage behavior change.

## 5 Future Work

Since the software is undergoing continual improvement, we would like to test the newest version which includes the ability to use Google Account linking for a better user experience. Linking to a Google account provides the ability to start the conversation and gain access to the data using different variations of prompts such as "How much more do I need to drink today?"

The ability to add different types of drinks could be beneficial as well to be able to manage a user's daily intake on more levels and possibly help with different types of med-ical conditions. This would require changes to the database structures, but can be scaled to include metrics and data to improve different user routines in different areas of life.

Another useful feature would be to add notifications that would nudge the user and remind them about different tasks that they could do as an intervention based on their data. Working with many users requires the automation of data analysis and export; this is an important step in the process to be able to fully automate the study and continually be able to monitor and perform changes over a period of time.

Some interventions that we would like to work on include a changing goal system that calculates the optimal goal based on previous patient data that is loaded when a user first enters the study and creates an estimate of how much a user should drink or sleep on any given day. In the latest software version, the users have a fixed goal that is intended to be met daily. Ideally, this would be customized to the patient's condition, and can be linked to health monitoring devices to provide more accurate insights. For example, a kidney disease patient may have dialysis appointments monitored and our Google Action would change the fluid requirements to appropriately prepare for the dialysis session.

There is an open research question with regards to the optimal method for nudging users towards survey adherence. By providing notifications in the correct situational context, we can improve the rate at which data is collected and thereby improve accuracy for follow-on interventions. Research shows a clear benefit to "just-in-time" and context appropriate user-friendly interventions for providing behavioral support at key times when an individual has the opportunity to change and is receptive to such support [10]. Google Assistant includes a built-in "Routines" function, which allows users to schedule notifications for completing necessary actions. This is perhaps the simplest solution, and is the implementation that we used for Alpha users of FluidMonitor and Sleepy. Users can even engage with their surveys by clicking on the notification that appears via Google Routine on their smart device. Alternative methods include scheduling the Google Actions with Assistant's alarm clock feature or establishing separate reminder notifications in the user's smartphone. Finally, although yet to be explored, our Google Actions could be automatically invoked at the conclusion of any other user interaction with their smart speaker; for example, after asking Google for the weather report, the VCA could immediately prompt the user to complete their sleep diary for the day.

One advantage of using COTS smart devices is the ability to integrate with other smart devices. Fitbit activity monitors can provide a wealth of knowledge about exercise and sleep patterns. Sleep interventions in Sleepy could be designed around smart-home controls, which can be used to control lighting and thermostat settings. FluidMonitor and Sleepy could benefit from sharing features with these technologies.

Finally, we recommend the implementation of a full-scale randomized study to discover and measure our system's true effect on behavioral change. Although studies of this nature exist in the literature, we suggest further study for the effects of our specific VCA approach on patients with fluid requirements or sleep disorders.

## 6 Limitations

Some limitations we found included the scaleability of the current system to accommodate a bigger trial, and the automation of the data analysis and processing. The architecture must be able to handle the amount of data automatically to increase the number of users. The Google Actions Console is relatively new, and thus documentation is still being created.

Error handling is a problem unique to the VCA approach. This issue was noticed in our user testing for FluidMonitor and Sleepy. When the software fails to accurately recognize the user's intended response, it is challenging for the user to identify the mistake and make a correction. Given a limited time window to respond to each survey question, users are prone to make speech errors or have their response misunderstood. Furthermore, without visual or auditory feedback to survey responses, the user may not notice an incorrectly captured response. This compromises the accuracy of data and future work is required to ensure that user responses are correctly stored. One solution is for the VCA to provide a "read-back" confirmation of the user's response, after which a correction can be made.

## 7 Conclusions

In this paper we validated the feasibility of deploying two VCAs for use cases that help individuals monitor their behav-ior related to fluid intake and sleep. We had the system tested, and demonstrated its utility in data storage and analysis with Alpha users. We created a study using VCAs to administer surveys and interventions based on our design and identified future directions and research areas. We also discussed the optimal ways to deliver those surveys and interventions. We introduced FluidMonitor, a Google Action with the purpose of measuring general fluid intake and health ratings to provide interventions. We also introduced Sleepy, with the purpose of providing a voice-assisted sleep diary and interventions to improve sleep. For the next step we propose the design of a long-term randomized controlled trial for Fluid Monitor and Sleepy to determine the extent to which VCAs can change behavior and improve quality of life for patients.

## Acknowledgments

The authors would like to thank Professor Mehdi Boukhechba for mentorship and guidance during this research.

## References

[1] B. Kinsella and A. Mutchler. Voice assistant consumer adoption in healthcare. [Online]. Available: https://voicebot.ai/wp-content/uploads/2019/10/voice\_assistant\_consumer\_adoption\_in\_healthcare\_report\_voicebot.pdf

[2] B. Kinsella. Nearly 90 million u.s. adults have smart speakers, adoption now exceeds one-third of consumers. [Online]. Available: https://voicebot.ai/2020/04/28/nearly-90-million-u-s-adults-have-smart-speakers-adoption-now-exceeds-one-third-of-consumers/

[3] Water: How much should you drink every day? - mayo clinic. [Online]. Available: https://www.mayoclinic.org/healthy-lifestyle/nutrition-and-healthy-eating/in-depth/water/art-20044256

[4] Fluid overload in a dialysis patient. [Online]. Available: https://www.kidney.org/atoz/content/fluid-overload-dialysis-patient

[5] C. Bérubé, T. Schachner, R. Keller, E. Fleisch, F. v Wangenheim, F. Barata, and T. Kowatsch, "Voice-based conversational agents for the prevention and management of chronic and mental health conditions: Systematic literature review," vol. 23, no. 3, p. e25933. [Online]. Available: https://www.jmir.org/2021/3/e25933

[6] S. Greuter, S. Balandin, and J. Watson, "Social games are fun: Exploring social interactions on smart speaker platforms for people with disabilities," in Extended Abstracts of the Annual Symposium on Computer-Human Interaction in Play Companion Extended Abstracts. ACM, pp. 429–435. [Online]. Available: https://dl.acm.org/doi/10.1145/3341215.3356308

[7] A. Cheng, V. Raghavaraju, J. Kanugo, Y. P. Handrianto, and Y. Shang, "Development and evaluation of a healthy coping voice interface application using the google home for elderly patients with type 2 diabetes," in 2018 15th IEEE Annual Consumer Communications Networking Conference (CCNC), pp. 1–5, ISSN: 2331-9860.

[8] J. Ooster, M. Krueger, J.-H. Bach, K. C. Wagener, B. Kollmeier, and B. T. Meyer, "Speech audiometry at home: Automated listening tests via smart speakers with normal-hearing and hearing-impaired listeners," vol. 24, p. 2331216520970011, publisher: SAGE Publications Inc. [Online]. Available: https://doi.org/10.1177/2331216520970011

[9] R. Maharjan, D. A. Rohani, P. Bækgaard, J. Bardram, and K. Doherty, "Can we talk? design implications for the questionnaire-driven self-report of health and wellbeing via conversational agent," in CUI 2021 - 3rd Conference on Conversational User Interfaces. ACM, pp. 1–11. [Online]. Available: https://dl.acm.org/doi/10.1145/3469595.3469600

[10] J. C. Walsh and J. M. Groarke, "Integrating behavioral science with mobile (mHealth) technology to optimize health behavior change interventions," vol. 24, no. 1, pp. 38–48. [Online]. Available: https://econtent.hogrefe.com/doi/10.1027/1016-9040/a000351

[11] S. Michie, L. Yardley, R. West, K. Patrick, and F. Greaves, "Developing and Evaluating Digital Interventions to Promote Behavior Change in Health and Health Care: Recommendations Resulting From an International Workshop," vol. 19, no. 6, p. e232. [Online]. Available: http://www.jmir.org/2017/6/e232/

[12] L. M. Ritterband, F. P. Thorndike, L. A. Gonder-Frederick, J. C. Magee, E. T. Bailey, D. K. Saylor, and C. M. Morin, "Efficacy of an internet-based behavioral intervention for adults with insomnia," vol. 66, no. 7, p. 8.

[13] L. M. Ritterband, F. P. Thorndike, K. S. Ingersoll, H. R. Lord, L. Gonder-Frederick, C. Frederick, M. S. Quigg, W. F. Cohn, and C. M. Morin, "Effect of a web-based cognitive behavior therapy for insomnia intervention with 1-year follow-up: A randomized clinical trial," vol. 74, no. 1, p. 68. [Online]. Available: http://archpsyc.jamanetwork.com/article.aspx?doi=10.1001/jamapsychiatry.2016.3249

[14] K. S. Hone and R. Graham, "Towards a tool for the subjective assessment of speech system interfaces (SASSI). natural language engineering 6(3/4):287–305."

[15] S. Hooker, A. Punjabi, K. Justesen, L. Boyle, and M. D. Sherman, "Encouraging health behavior change: Eight evidence-based strategies," vol. 25, no. 2, pp. 31–36. [Online]. Available: https://www.aafp.org/fpm/2018/0300/p31.html

[16] P. Karppinen, H. Oinas-Kukkonen, T. Alahäivälä, T. Jokelainen, A.-M. Teeriniemi, T. Salonurmi, and M. J. Savolainen, "Opportunities and challenges of behavior change support systems for enhancing habit formation: A qualitative study," vol. 84, pp. 82–92. [Online]. Available: https://linkinghub.elsevier.com/retrieve/pii/S1532046418301187

Appendix A
SHUTi Sleep Diary Questionnaire
SHUTi Sleep Diary Questionnaire provided by Professor Mehdi Boukhechba

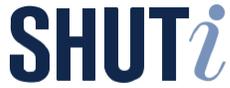

# SLEEP DIARY

**SHUTi accepts Sleep Diaries for only the past 3 nights. Be sure to sign-in often to record your sleep!**

| Date (the morning you woke up) | Example 6/18/18 | | | |
|---|---|---|---|---|
| 1. What time did you get into bed? | 10:15pm | | | |
| 2. What time did you try to go to sleep? | 11:30pm | | | |
| 3. How long did it take you to fall asleep? | 1 hr 15 min | | | |
| 4. How many times did you wake up, not counting your final awakening? In total, how long did these awakenings last? | 3 times<br>1 hr 10 min | | | |
| 5. What time was your final awakening? | 6:00am | | | |
| 6. What time did you get out of bed for the day? | 6:30am | | | |
| 7. How would you rate the quality of your sleep?<br>Very poor \| Poor \| Fair \| Good \| Very good | Poor | | | |
| 8. In total, how long did you nap or doze, yesterday? | 40 min | | | |
| 9. How many drinks containing alcohol did you have? What time was your last beverage with alcohol? | 2<br>8:00 pm | | | |
| 10. Did you take any over-the-counter or prescription medications to help you sleep? If so, list medication(s). | Melatonin, 1mg @ 9:00pm<br>Ambien, 5mg @ 10:00pm | | | |
| My Personal Notes<br>This is your space to record whatever you wish to track or remember about your sleep. | Poor sleep due to a cold | | | |